\documentclass[12pt,preprint]{aastex}

\def\spose#1{\hbox to 0pt{#1\hss}}
\def\approxlt{\mathrel{\spose{\lower 3pt\hbox{$\sim$}}
        \raise 2.0pt\hbox{$<$}}}
\def\approxgt{\mathrel{\spose{\lower 3pt\hbox{$\sim$}}
        \raise 2.0pt\hbox{$>$}}}

\def\multleft#1{\hbox to size{\vbox {\halign {\lft{##}\cr #1}}\hfill}\par}
\def\multright#1{\hbox to size{\vbox {\halign {\rt{##}\cr #1}}\hfill}\par}

\def\boxit#1{\vbox{\hrule\hbox{\vrule\kern3pt\vbox{\kern3pt
          #1 \kern3pt}\kern3pt\vrule}\hrule}}

\def\cm{{\rm\thinspace cm}}

\def\erg{{\rm\thinspace erg}}

\def\km{{\rm\thinspace km}}

\def\s{{\rm\thinspace s}}

\def\chisq{\hbox{$\chi^2$}}

\def\cmsq{\hbox{$\cm^2\,$}}
\def\pcmsq{\hbox{$\cm^{-2},$}}

\def\ergpcmsqps{\hbox{$\erg\cm^{-2}\s^{-1}\,$}}

\def\ergps{\hbox{$\erg\s^{-1}\,$}}

\def\kmps{\hbox{$\km\s^{-1}\,$}}

\def\pcmsq{\hbox{$\cm^{-2}\,$}}

\received{}
\accepted{}
\slugcomment{submitted to ApJ} 
\shorttitle{Suzaku Observation of The Circinus Galaxy}
\shortauthors{Yang et al.}

\begin{document}
\title{Suzaku Observations of the Circinus galaxy}

\author{Y. Yang\altaffilmark{1,2},
A. S. Wilson\altaffilmark{1},
G. Matt\altaffilmark{3},
Y. Terashima\altaffilmark{4},
L. J. Greenhill\altaffilmark{5}} 
\altaffiltext{1}{Department of Astronomy, University of Maryland, College Park, MD,  20742, USA}
\altaffiltext{2}{Current Address: Department of Astronomy, University of Illinois, Urbana-Champaign, IL, 61801} 
\altaffiltext{3}{Dipartimento di Fisica, Universita degli Studi ``Roma tre", via della Vasca Navale 84, I-00046, Roma, Italy}
\altaffiltext{4}{Department of Physics, Faculty of Science, Ehime University, Matsuyama, 790-8577, Japan}
\altaffiltext{5}{Smithsonian Astronomical Observatory, Center for Astrophysics, MS42; 60 Garden St; Cambridge, MA 02138 USA}

\begin{abstract}
We report {\it Suzaku} observations of the active, Compton-thick Circinus galaxy. Observations
were obtained with both the X-ray Imaging spectrometer (XIS) and the Hard X-ray Detector (HXD). Below 10 keV, the nuclear spectrum
is dominated by radiation reflected from cold dense gas of high column density, while above 13~keV the radiation 
is directly transmitted nuclear emission seen through a column density of $\simeq 4 \times 10^{24}$~\pcmsq. In the 0.2--10~keV
band, the XIS spectrum is contaminated at 5\% level by the brightest off-nuclear source in Circinus (CG X-1), but drops to 
1\% in the 5-10~keV and is negligible at higher energies. We find no significant evidence 
for variability in the hard ($>12$~keV) emission. The Circinus is marginally detected with the HXD/GSO in 
the 50--100~keV band at 2.5$\sigma$ level. We model the 3-70~keV band XIS+PIN spectra with a four components: the Compton transmitted nuclear emission, the reflected nuclear emission, a soft power law (representing a combination of scattered nuclear emission, extended emission and contamination by sources in the galaxy below a few keV). The hard nuclear power-law is found to have a photon index $\Gamma_h \simeq 1.6$, very similar to the soft power-law. The high energy cut-off is $E_C \simeq 49$~keV. These results agree with those from BeppoSax. An extrapolation of this model up to the GSO band shows good agreement with the GSO spectrum and supports our detection of the Circinus up to $\simeq 100$~keV.      
\end{abstract}


\section{Introduction}
The Circinus galaxy is a large, nearby ($4\pm 1$~Mpc; \markcite{freeman77}{Freeman} {et~al.} 1977) galaxy that harbors both a circumnuclear star burst and a Seyfert 2 nucleus.  Evidence for an obscured Seyfert 1 nucleus is provided by the finding of a  broad (FWHM $\simeq$ 3000 \kmps)  H$\alpha$ line component in polarized light \markcite{oliva98}({Oliva} {et~al.} 1998). The picture of an obscured Seyfert 1 nucleus is also supported by both the discovery of highly ionized gas extending along the minor axis of the galaxy, with a morphology that is reminiscent of the ionization cones seen in other Seyfert galaxies \markcite{marconi94, wilson00}({Marconi} {et~al.} 1994; {Wilson} {et~al.} 2000), and direct X-ray detection of the nucleus through a column density of $\sim 4 \times 10^{24}$~\pcmsq (\markcite{matt99}{Matt} {et~al.} 1999, hereafter M99; \markcite{soldi05}{Soldi} {et~al.} 2005).  

{\it ASCA}, {\it BeppoSax} and {\it XMM-Newton}  spectra  below 10 keV show both a flat (hard) spectrum and high equivalent width Fe K$\alpha$ emission, characteristic of Compton reflection from cold gas illuminated by a power-law continuum. A single scattering Compton shoulder was found in {\it Chandra} HETG observations \markcite{bianchi02}({Bianchi} {et~al.} 2002) and was confirmed by {\it XMM-Newton} \markcite{molendi03}({Molendi}, {Bianchi}, \& {Matt} 2003), demonstrating conclusively the existence of Compton thick matter.   Because of their very flat hard X-ray spectra, Compton-thick AGNs are a key ingredient in the synthetic model of the hard cosmic X-ray background \markcite{setti89, madau94}({Setti} \& {Woltjer} 1989; {Madau}, {Ghisellini}, \& {Fabian} 1994), which peaks at 30--40~keV and rolls over at higher energies. After  NGC~4945 and  NGC~1068, Circinus galaxy is the third brightest Compton-thick AGN that allows detailed study.    

In this paper, we present a broad band observation of the Circinus galaxy using the {\it Suzaku} telescope. The low background and high sensitivity of {\it Suzaku} allow improved  constraints on the nuclear continuum and the nature of the absorbing and reflecting matter. In \S~\ref{data}, we describe the observation and data reduction. The results are presented in \S~\ref{results}.  We discuss the results and their implications in \S~\ref{concl}.

\section{The {\it Suzaku} Observation and Data Analysis
\label{data}}
The Circinus galaxy was observed for $\sim 100$~ks with {\it Suzaku} on July 21, 2006. The data were obtained with the four X-ray Imaging Spectrometer (XIS,\markcite{koyama07}{Koyama} {et~al.} 2007) CCDs, and the non-imaging, collimated hard X-ray detector (HXD).  Each of the four XIS CCDs (XIS0, 1, 2 and 3) is located in the focal plane of a foil X-ray telescope \markcite{serlemitsos07}({Serlemitsos} {et~al.} 2007). The CCD chip for each camera has a dimension of 1024 $\times $ 1024 picture elements (``pixels''), and covers a $17.8' \times17.8'$ region on the sky. The CCD cameras are sensitive in the range 0.2--12~keV. One of the XISs, XIS1, uses a back-side illuminated (BI) CCD, while the other three use front-side illuminated (FI) CCDs. The BI CCD has better quantum detection efficiency at sub keV energies. Each CCD sensor has two $^{55}$Fe calibration sources located on the side wall of the housing; these sources illuminate two adjacent corners of the CCD. The spectral resolution of the XIS at 6 keV is 130 keV.  The half power diameter of the X-ray telescope is $2\arcmin$. The HXD uses a novel well-type phoswich counter (Takahashi et al. 2006) which greatly reduces the instrument background. The HXD sensor consists of 16 main detectors and surrounding 20 crystal scintillators for active shielding.  Each unit  consists of two types of detectors: a GSO scintillation counter, and 2 mm-thick PIN silicon diodes located inside the well and in front of the GSO scintillator. The PIN diodes are most sensitive below $\sim 60$ keV, while the GSO  is sensitive above $\sim 40$ keV.  Our observations were performed during the period when bias voltages for 16 out of 64 PIN diodes were reduced from 500 V to 400 V to suppress the rapid increase of noise events that possibly result from in-orbit radiation damage. Since the thickness of the depletion layer depends on the bias, this change slightly affected the energy response of the 16 PIN diodes especially in the energy band higher than 20~keV.  

The observation started on July 21, 2006 at 12:29:57 UT, and stopped on July 23, 2006 at 1:08:24 UT. The nominal point of the observation lying at the center of the XIS field, is at $\alpha_{2000} = 14^{h}13^{m}09^{s}.84$, $\delta_{2000} = -63\arcdeg 20\arcmin 24\arcsec$. The data were first reprocessed with version 1.2.2.3 calibrations and analyzed with FTOOLS included in HEADAS 6.2. Preliminary results were presented in \markcite{yang07}({Yang} {et~al.} 2007). After we finished the first draft of the paper, version 2 reprocessed data  became available. Improvements of the XIS calibration have resolved some of the issues in the XIS spectra we found when using version 1 processing. We therefore use the version 2 reprocessed XIS data (process version 2.0.6.13). For the HXD data, we found no significant difference in our PIN spectra using either version of data. Since the reproductivity of version 2 processed HXD/GSO background is still unavailable, which affects our GSO results, we opt to keep using version 1 processed HXD data, where the systematics are better known. Following the recommendation of the {\it Suzaku} instrument team, we selected only the 48 PIN diodes that used a bias voltage of 500 V.   The cleaned exposure times for XIS and HXD are 108.3~ks and 86.0~ks, respectively.  

\begin{figure}
\includegraphics[scale=0.8,angle=0]{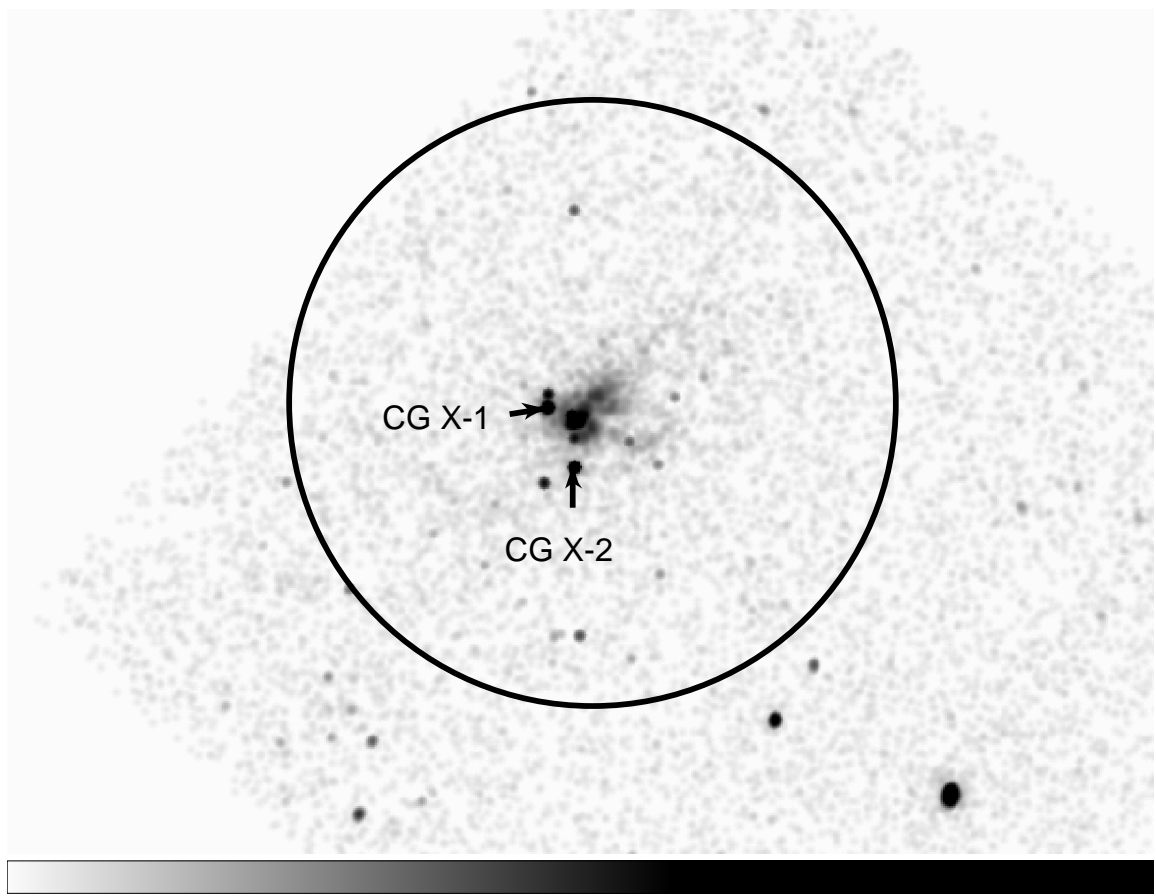}
\caption[image]{ {\it Chandra} 0.5-7 keV ACIS-S image of the Circinus galaxy (Obs Id 356, PI: Andrew Wilson). The image is slightly smoothed for better viewing. The $2.5\arcmin$ radius circular XIS spectral extraction region is shown. The two brightest ULXs included in the extraction region, CG X-1, and X-2 in Bauer et al. (2001), are marked. A bright point source $\sim 5\arcmin$ SW of the Circinus galaxy is also seen in the {\it Chandra} image. This source is within the field-of-view of the HXD detector.  
\label{xis0image}}
\end{figure}

We extracted the XIS spectra and light curves  from a $2.5\arcmin$ radius region centered on the Circinus galaxy. The background spectra were extracted from a set of circular regions away from the Circinus galaxy and other bright point sources and extended emissions. The spectra from the calibration sources were also extracted. The broad point spread function of the XRTs mixes the emission from the nucleus with the emission from several point sources, including the two bright ultra-luminous X-ray sources (ULXs), named CG X-1, CG X-2  in Bauer et al. (2001), and that from the diffuse gas seen in {\it Chandra} images \markcite{smith01}({Smith} \& {Wilson} 2001).  This may complicate the interpretation of the soft X-ray spectrum.  We show the XIS extract region overlaid on the {\it Chandra} image of the Circinus galaxy field in Fig.~\ref{xis0image} to illustrate the location of the ULXs.

Below 100~keV, the passive collimator of the HXD defines a $34\arcmin \times 34\arcmin$ field of view.  Instead of rocking between source and blank sky, the non-cosmic background of the HXD is modeled  based on the earth occultation data, monitoring data of the actual particle flux, and the orbit data of the satellite. The accuracy of the background is determined by the reproductivity of the model rather than the statistical error of the background counts. Since the non X-ray background contributes significantly to the HXD counts, the accuracy of the model background is the major factor determining the sensitivity of the instrument. The accuracy of the model background can, in most cases, be tested by comparing the simulated background with that obtained during Earth occultation. This, however, is not the case for the Circinus galaxy because the galaxy is located in the continuous viewing zone of {\it Suzaku} with earth elevation angle always $>-5\arcdeg$.  We therefore rely solely on the HXD non X-ray background files provided by the HXD team in our analysis. The uncertainty of the background can only be stated in a statistical sense. 

\section{Results
\label{results}}
\subsection{Variability}
\subsubsection{XIS}
\begin{figure}
\includegraphics[scale=0.6,angle=270]{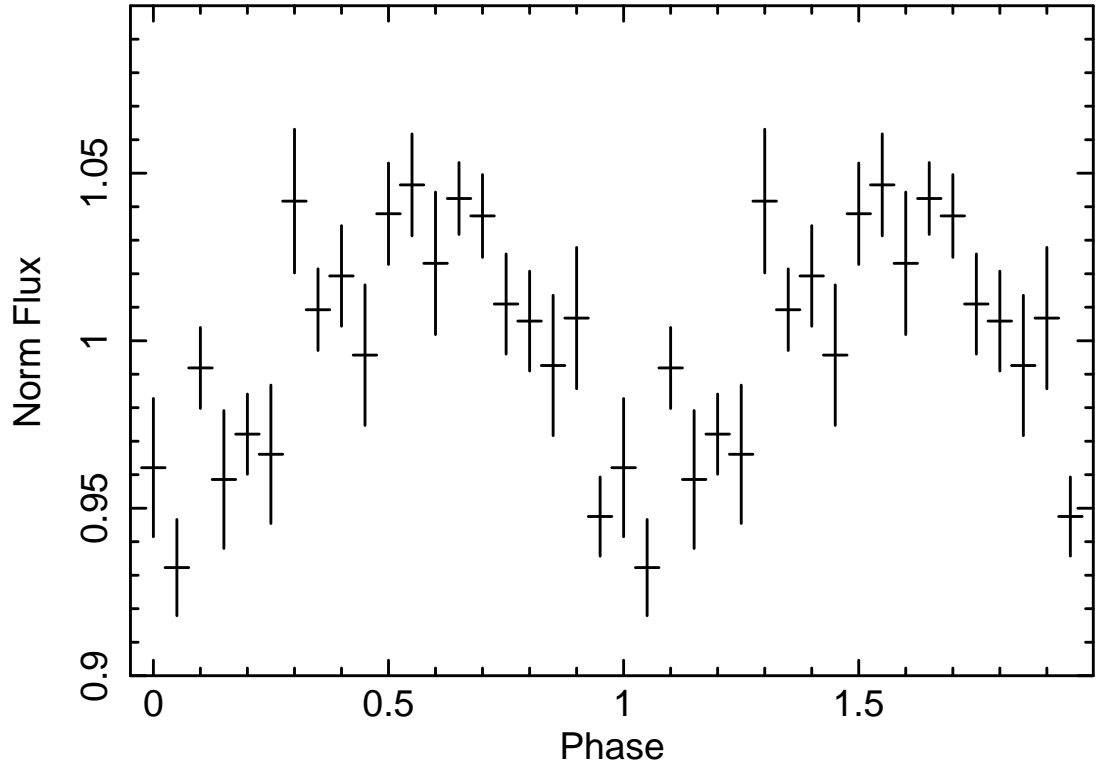}
\caption[image]{ The folded 0.2-10 keV XIS light curve of the Circinus Galaxy. Light curves from all of the four XIS instruments are included. The folding period is 27~ks and the count rate is normalized to the mean value of 2.25~cts/s.  
\label{xis0ltcv}}
\end{figure}

The brightest ($L_x = 3.71 \times 10^{39}$~\ergps in the 0.5 to 10 keV band, corrected for absorption) off nuclear source in Circinus -- CG X-1 -- was found to be periodic with period 27.0~ks through {\it Chandra} observations by Bauer et al. (2001). We examine the folded 0.2--10~keV band XIS light curve (Fig.~\ref{xis0ltcv}) and the period is clearly detected. The variation amplitude between minimum and maximum is 10\%. However, the amplitude reduces to only 2\% in the 5-10~keV band, indicating the spectrum of the variable source should be steeper than that of the nuclear source. This result agrees with the notion that the spectrum in this band is dominated by the reflected component of the nuclear emission.  
\subsubsection{PIN}
On the other hand, the observed 12-50~keV PIN light curve suggests a slow variability on a time scale of $\sim 50$~ks during our observation, even after the time variable deadtime correction has been performed. The standard deviation of the count rate is 15\% (4 ks binning).   However, the model background count rate is $\sim 60\%$ of the total count rate and thus can be an important source of the observed variability. For 4~ks bins, the uncertainty of the PIN non X-ray background reproductivity measured by the standard deviation of the residual between the earth occultation and blank sky data and the background model is  $\sigma \sim 7\%$ \markcite{mizuno06}({Mizuno} {et~al.} 2006, Fig.~3). The residual also tends to be larger during the SAA passages (because the background is harder to model during these periods), and produces a slow variation on time scales $\lesssim 70$~ks. This level of variability translates to a $\sigma \sim 10\%$ uncertainty in the source count rate, which is about the same level of variation in our PIN light curve.  Therefore, our detected variability in the hard X-ray emission is probably not real.  This conclusion is consistent with the emission being Compton scattered in a thick torus and the intrinsic variability has been suppressed and smoothed by the scattering medium.
 The lack of variability in Circinus galaxy compares interestingly to NGC 4945, which show strong variability in hard X-ray (Iwasawa et al, 1993; Madjeski et al. 2000 and Itoh et al. 2008). NGC 4945 has very similar spectral properties as Circinus galaxy with $N_H \sim 5 \times 10^{24}$~\cmsq (Itoh et al. 2008). Using Monte Carlo simulations, Madjeski et al. (2000) show that the large variability is consistent with the absorber in NGC 4945 being a geometrically thin disk. Since Circinus galaxy has similar optical depth as NGC 4945, we can estimate the geometry of the torus by applying the same simulation result shown in Fig.3 of Madjeski et al. (2000). Assuming a Fe abundance of unity, our upper limit of variability ($< 15\%$) suggests the half angle subtended by the torus viewed at the central source $> 80 \arcdeg$. This is roughly consistent with the estimate subtended of the opening angle of the torus in Circinus galaxy by Ghisellini et al. (1994).  


\subsection{GSO detection}
\begin{figure}
\includegraphics[scale=0.57,angle=270]{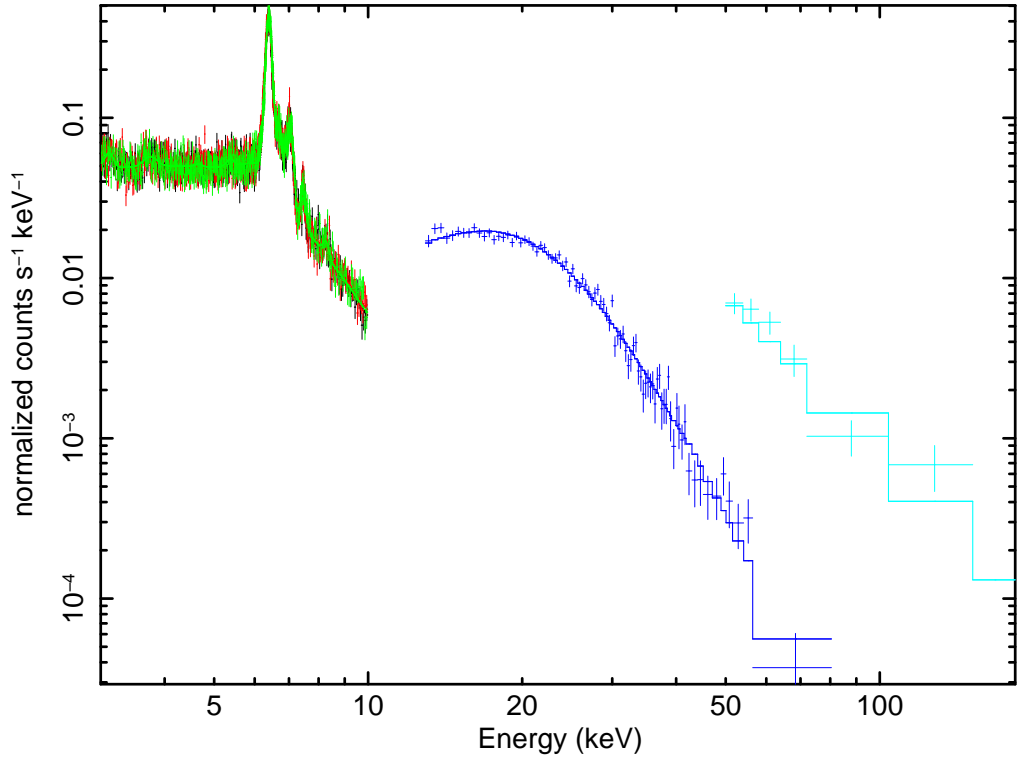}
\caption[spec]{The 3-200~keV {\it Suzaku} spectrum of Circinus galaxy and the folded best-fit model (model 1 in the text). The color representation for the spectra from each of the instruments is defined as: black--XIS0, red--XIS2, green--XIS3, blue--HXD/PIN, and cyan--HXD/GSO. Note that while the GSO spectrum agree very well with the extrapolation of the best-fit XIS+PIN model, it is not included in the spectral models. 
\label{spec}}
\end{figure}

The background subtracted GSO count rate for the Circinus galaxy in the 50--100 keV band is 0.11 cts/s. The corresponding count rate of the background is 5.43 cts/s. The uncertainty of the background model for a 1 day exposure in the 50-100~keV band is $\sim 0.81\%$ \markcite{takahashi06}({Takahashi} {et~al.} 2007, Fig.~3). Therefore, the Circinus galaxy is detected in this band at a $\sim 2.5\sigma$ level. In Fig.~\ref{spec} we show the extrapolation of the 3--70~keV XIS+PIN spectra to the GSO band up to 200 keV using the spectral model described in \S~\ref{joint}. A good agreement is found between the data and the extrapolation of the model. Better constraints on the GSO spectrum await the improved accuracy of the GSO background. 

\subsection{The Spectra
\label{spectra}}
In Fig.~\ref{spec} we show the {\it Suzaku} spectrum in the 3--200~keV band. Before modeling the broad band spectrum of Circinus Galaxy,  it is instructive to first analyze the XIS and HXD spectra separately. This allows us to understand the calibration of the instruments and the contamination from off nuclear sources in each instrument.  Below 10~keV,  the spectrum is dominated by reflected nuclear emission, while above 20~keV, the spectrum is almost completely Compton transmitted emission.  Analyzing the XIS and HXD spectra separately make it easier to understand  each component of the spectrum without too much confusion from other components. Degeneracies between parameters such as the intrinsic photon index and the absorption column density are inevitable when using only the XIS or PIN spectrum,  and are best addressed with  joint analysis using both XIS and HXD spectra, which is discussed in \S~\ref{joint}. 

\subsubsection{The X-ray Imaging Spectrometer Spectra
\label{xis_spec}}
Within the $2.5\arcmin$ extraction radius, the most likely contamination of the XIS spectrum comes from the two ULXs in the field \markcite{bauer01, smith01}({Bauer} {et~al.} 2001; {Smith} \& {Wilson} 2001).  Without simultaneous {\it Chandra} observations, it is hard to estimate the true contribution from these off-nuclear sources. In most of the previous observations, the 2-10 keV fluxes of the two ULXs remain about an order of magnitude lower than that of the nucleus. However, on March 14, 2001, the {\it Chandra} flux from CG X-1 reached the same level of the nuclear flux \markcite{smith01, bianchi02}({Smith} \& {Wilson} 2001; {Bianchi} {et~al.} 2002). We assume that the contamination from the ULXs and other unresolved soft emissions can be represented with a single power-law. This component dominates the soft X-ray emission below a few keV, and we call it the soft power-law component in the following discussions. By modeling the XIS spectrum with a reflection component from dense cold gas near the blackhole (pexrav model in XSPEC, \markcite{magdziarz95}{Magdziarz} \& {Zdziarski} 1995) plus the soft power-law with {\it ad hoc} added emission lines, we found the total 2--10~keV flux within the extraction region to be $1.73 \times 10^{-11}$~\ergpcmsqps. This flux lies between that of the two {\it BeppoSax} observations on Mar. 13, 1998 and Jan. 7, 2001. The soft power-law component is found to have $\Gamma_s = 1.54$ and $N_{H} = 3.8 \times 10^{21}$~\pcmsq, where the absorption column density includes both that from the Circinus galaxy and the our own Galaxy. The 2-10~keV flux of the component is $5.76 \times 10^{-12}$~\ergpcmsqps. This is consistent with the best-fit soft power-law in M99, but higher than the average flux of either CG X-1 ($9 \times 10^{-13}$~\ergpcmsqps)  or CG X-2 ($1.3 \times 10^{-12}$~\ergpcmsqps).  This suggests that 1) the ULXs are likely to be in states similar to those during the March 1998 observation, and 2) the soft emission may come mostly from the scattered nuclear emission rather than from the ULXs. Moreover, since the photon index of the soft power-law is higher than that of the cold reflection component in our model, the relative contamination from the ULXs should drop significantly at higher energies. In the 5--10~keV band, the soft power-law component contributes only $\sim 20\%$ to the total flux. The very low amplitude of the 27~ks periodical variation at high energies also suggests that the spectrum of CG X-1 is steep in this observation.  Nevertheless, the contribution from the ULXs may still be significant sources of contamination in the energies below 2 keV.  For this reason, we focus our analysis on the hard X-ray spectrum $> 3 $~keV,  and defer the detailed study of the soft X-ray emission to a future paper. 

We compare the XIS energy scales by fitting the strongest emission lines in the 3--10~keV spectra.  Small, but significant, differences are seen in the line energies between different XIS spectra. The largest difference is found between the line energies using the BI XIS-1 spectrum and those using the  FI CCD spectra. The energies of lines in the XIS-1 spectrum are higher than those from the FI CCDs. For example,  the Fe K$\alpha$ line in the XIS-1 spectrum is $\sim 25$ eV higher than the best-fit energy of the line using the rest of the XIS detectors. This difference is significantly larger than the nominal 0.2\% uncertainty level of the energy calibration. On the other hand, such discrepancy is not seen in the emission lines from the calibration sources located at the corners of the CCDs. This problem has been seen in observations taken after mid-2006 in observations without spaced-row charge injection (Hamaguchi 2007, private communication). A possible cause is that the charge transfer inefficiency is not properly corrected in these observations. The best-fit Fe K$\alpha$ line rest frame energy using the FI CCDs is 6.401~keV, which is very close to the theoretical value of the doublet (6.391~keV and 6.405~keV). This is also the case for the rest of the emission lines in the 3--10~keV spectrum, indicating the energy calibration of the FI CCDs in our observation are probably better than that of the BI CCD. Given that the effective area of XIS-1 is significantly lower in the 3--10~keV band than the FI CCDs, dropping the data from XIS-1 does not significantly change the signal-to-noise, but significantly improves the statistics of the fit. Thus, we only analyze the spectra from the FI CCDs in this paper. 

The model we employed to fit the XIS spectra  consists of three components:  the reflected nuclear emission (pexrav in XSPEC), a soft power-law that represents the scattered nuclear emission and the emission from contamination sources, and a set of emission lines. To reduce the number of free parameters, we have adopted in the pexrav model the inclination angle of $\cos(i) = 0.45$ and set the reflection parameter to -1 to make the component pure reflection.
Since the normalization and the inclination of the slab $i$ is highly degenerate, choosing  $\cos(i) = 0.45$ (or $i \sim 63\arcdeg$) is mainly to reduce the number of free parameters. For a cylindrical torus, the reflecting surface is the inner side of torus which is perpendicular to the accretion disc. Therefore, choosing $i = 63\arcdeg$ means we are looking at the torus at $27~\arcdeg$. This choice is consistent with the estimates that the inclination of the system is $ < 40\arcdeg$ (Matt et al. 1999).   
 We also assume the reflecting gas to have the same metallicity as that of the Sun except for element Fe, which is left as a free parameter\footnote{We use the metallicity table from \markcite{anders89}{Anders} \& {Grevesse} (1989)}.  The upper cutoff energy of the spectrum of the reflected component is poorly constrained with only the XIS spectrum, and the value from the PIN spectrum $E_C = 48.7$~keV (see \S~\ref{pin_spec}) is adopted.  We consider the following two cases for the continuum: 1) The soft power-law parameters are frozen to those found in the 0.3--10~keV spectral fitting. This component can be a combination of emissions from off-nuclear sources and the scattered emission. The intrinsic photon index of the reflected component is found to be $1.79_{-0.13}^{+0.06}$;  2) If the emission above 3 keV is predominantly the nuclear emission scattered by ionized matter thus the photon index of the soft power-law and the intrinsic photon index of the reflection component should be the same (with the assumption that scattering dominates over the line emission, which is justified above 3~keV), and are thus linked in the spectral fitting. The photon index in this case is found to be $1.56_{-0.21}^{+0.18}$. 

\subsubsection{The Hard X-ray Detector spectrum
\label{pin_spec}}
Besides the two bright ULXs, CG X-1 and X-2, another bright point source located at $\sim 5\arcmin$ SW of the Circinus galaxy, is also within the field of view of the HXD and could be a source of contamination. The source can be seen in the full band XIS image, but not detected in the 10--13~keV image, suggesting the contribution from this source is weak compared to the nucleus of Circinus galaxy in the hard band and thus can be ignored in the HXD spectrum.  

We do not include the PIN data below 13~keV because of the considerable noise in this band, likely caused by thermal events. Since the Circinus galaxy is only $\sim 2.5\sigma$ above the GSO background in the 50--100~keV band, the uncertainty in the GSO background can introduce significant uncertainty in the spectrum. We therefore choose not to include the GSO data in the spectral modeling. This, however, limits our ability to constrain the high energy cut-off of the nuclear emission, which is likely to be $\sim 50$~keV. 

The cosmic background in the PIN band is estimated using the HEAO-1 best-fit spectrum \markcite{boldt87}({Boldt} 1987). 
$$
F(E) = 9.0 \times 10^{-9} \left( \frac{E}{3\thinspace {\rm keV}}\right) ^{-0.29} \exp \left(-\frac{E}{40 \thinspace {\rm keV}}\right)  {\rm \thinspace erg\thinspace cm^{-2}\thinspace s^{-1} \thinspace str^{-1} \thinspace keV^{-1}}
$$
and the response file for uniform emission in an area of $2\arcdeg \times 2\arcdeg$. We use XSPEC to simulate the CXB spectrum and subtract it from the PIN spectrum. The contribution of the CXB to the PIN count rate is $\sim 5.7\%$.  The Circinus galaxy lies $\sim 10\arcdeg$ away from the the bulk of the Galacic ridge emission \markcite{warwick85}({Warwick} {et~al.} 1985).  Using the spatial and spectral models of the Galactic ridge emission \markcite{yamasaki97, krivonos07}({Yamasaki} {et~al.} 1997; {Krivonos} {et~al.} 2007), we estimated the PIN count rate from the Galactic ridge emission to be $\lesssim 1\%$ and thus ignored. 

\begin{figure}
\includegraphics[scale=0.57,angle=270]{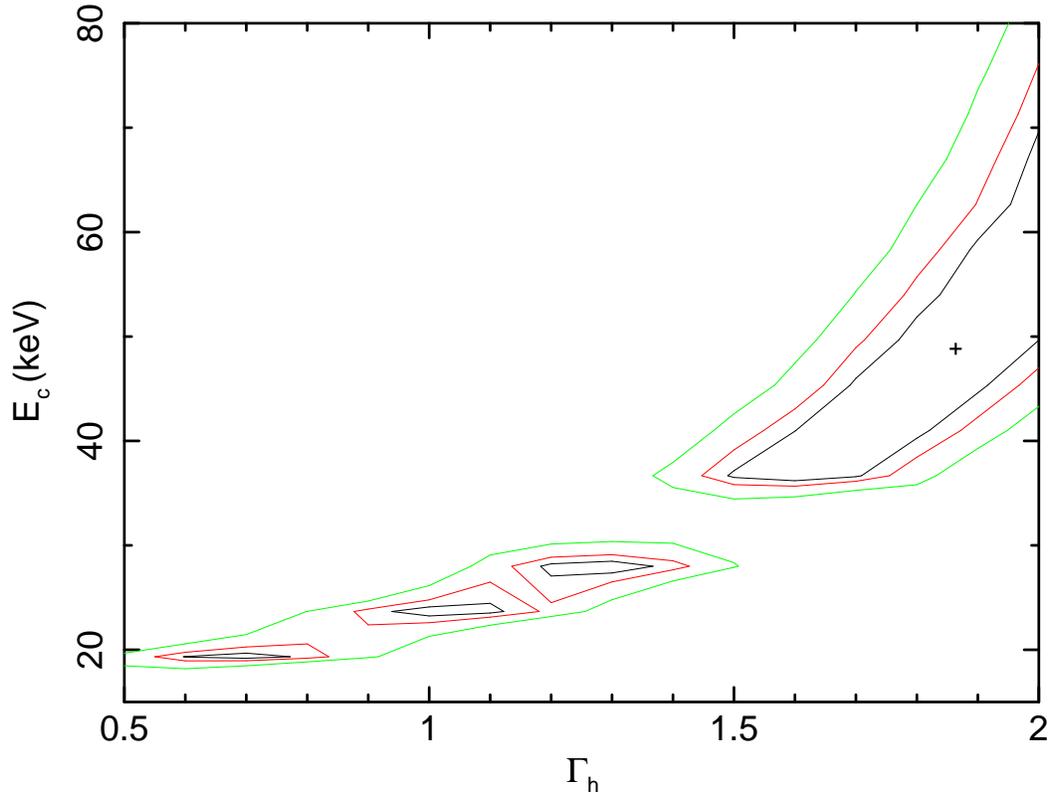}
\caption[Ec_cont]{$\chisq$ as  a function of $\Gamma_h$ and $E_C$ when the 20--70~keV PIN spectrum is modeled with a nuclear power-law viewed through a Compton transmitting medium (see text). The contours correspond to 1, 2 and 3-sigma confidence levels. 
\label{Ec_cont}}
\end{figure}

The PIN spectrum above 20~keV is well described by a heavily absorbed power-law with high energy cut-off, which probably originates from the nuclear emission transmitted through a torus or thick disk. We fit the 20--70~keV PIN spectrum with a Compton transmission model based on Monte Carlo simulations in a spherical distribution of matter (see M99 and references therein).  The model includes Compton down scattering and the decline of the Klein-Nishina cross section at high energies. A solution with photon index $\Gamma = 1.89_{-0.32-0.11}^{+0.20+0.28}$ and $E_C = 48.7_{-7.2-11.}^{+11.+49.}$~keV is found. The errors are  the statistical errors that correspond to the 90\% confidence ranges (the first numbers), and the systematic errors caused by the $\sim 4\%$ (90\% confidence) uncertainty of the PIN background (the second numbers). We estimate the systematic errors by varying the normalization of the PIN background by $\pm 4\%$. In the following, if only one set of errors are quoted, they correspond to only the statistical errors.  The photon index and the high energy cut-off  values agree very well with the {\it BeppoSax} results. The photon index also agrees with the best-fit XIS value.  However, solutions with flatter spectra and lower cutoff energies also exist.  In Fig.~\ref{Ec_cont}, we plot a contours map of $\chisq$ as a function of $\Gamma$ and $E_C$. The plot shows multiple local minima with overlapping $2\sigma$ contours. In fact,  some models with very flat spectra ($\Gamma <1$) produce slightly but insignificantly smaller $\chisq$ values.  This makes XSPEC converge to unphysical solutions when the parameter space is searched.  We therefore fix $E_C = 48.7$~keV when studying the confidence range of other spectral parameters (\S~\ref{joint}). The best-fit column density is found to be $N_H = 5.0_{-2.9-4.5}^{+3.9+3.6} \times 10^{24}$~\pcmsq, which agrees very well with previous results.  The uncertainty of $N_H$ is large because the parameter is best constrained in 10--20~keV band. However, the reflection component also contributes significantly in this band and thus $N_H$ is best determined by joint fit of the XIS and HXD spectra. 

\subsubsection{Overall 3--70 keV spectrum
\label{joint}}
We now study the broad band spectrum using both the XIS and HXD spectra. Again only the XIS spectra from 3--10~keV and the HXD/PIN spectrum between 13--70~keV were used in our analysis. Following M99,  the broad band emission is modeled by combining the XIS and PIN models we have described. In this model, the nucleus is obscured by a torus with a large covering factor, and only the reflection from the far side of the torus/disk is directly visible.  The model assumes the form
\begin{equation}
 F(E) = [A_T T(E, N_{H,T}, \Gamma_h, E_c)+A_R R(E, \Gamma_h,E_c)+A_s E^{-\Gamma_s}+lines] e^{-N_{H}\sigma_{ph}}, 
\end{equation}
where $T(E, N_{H,T}, \Gamma_h, E_c)$ is  the transmitted emission through an absorption column density $N_{H,T}$, $A_T$ is the normalization, $\Gamma_h$ is the photon index and $E_C$ the high energy cut-off;  $R(E, \Gamma_h,E_c)$ is the reflected nuclear emission by cold dense gas near the central engine or in the torus, $A_R$ is its normalization; $\Gamma_s$ is the photon index of the soft power-law, and $A_s$ its normalization; {\it lines} represents the flux of the emission lines; and $N_{H}$ is the column density of matter external to the nucleus (in the Circinus galaxy and our Galaxy). 

As in \S~\ref{xis_spec}, we consider two cases:  the photon-index of the soft power-law is fixed to the value from the 0.2--10~keV fit (model 1), and  the photon index of the soft power-law is tied to the intrinsic power-law index of the nuclear emission (model 2).  In model 1 the contribution of the soft power-law component to the 13--70 keV PIN count rate is $<5\%$.  The column density of matter external to the nucleus $N_{H}$ is also fixed to the best-fit XIS value. The high energy cutoff is fixed to 48.7~keV to avoid unphysical solutions with very flat spectra and low cutoff energies. We summarize the best-fit parameters in Table~\ref{tab_cont}. 

It is obvious that the two models are in fact very similar and the statistics are almost identical. Model 2 accounts naturally for the similarity between the soft and hard power-law components. Column densities of $\sim 4 -5 \times 10^{24}$ are found in both models, which also agree  with the previous findings from  {\it BeppoSax} and {\it Integral}. In model 1, we estimate the luminosity of the nucleus (corrected for absorption) in the 2-10~keV (2--200~keV) band to be $1.1 \times 10^{42}$~\ergps ( $2.4 \times 10^{42}$~\ergps) assuming a distance of 4~Mpc. 

We finally examine if the spectra can be modeled without the Compton transmitted component. In this case, the hard X-ray emission is purely 
reflected emission. We fit the 3--70~keV the spectra using the model in \S~\ref{xis_spec} (case 1). This model yields very flat intrinsic 
power-law with $\Gamma_h \sim 0.34$, and unusually large Fe abundance ($\sim 2.9$ solar). We consider these parameters unlikely for Circinus galaxy.       

\subsubsection{Emission lines and features
\label{lines}}
\begin{figure}
\includegraphics[scale=0.57,angle=270]{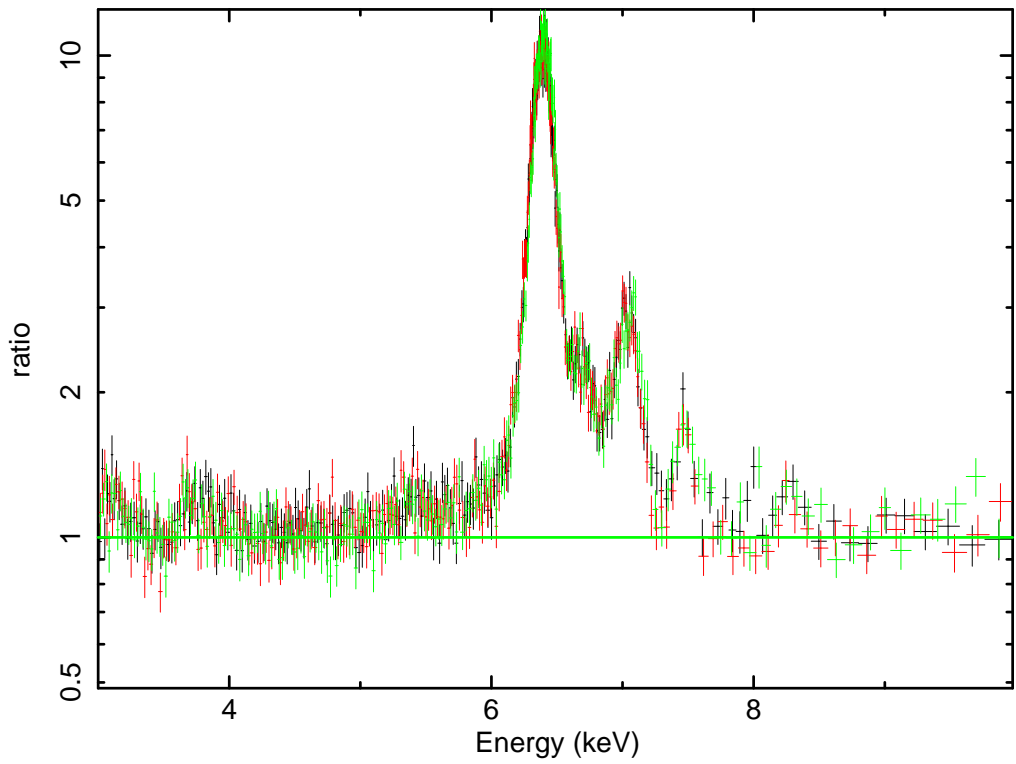}
\caption[models]{The ratio of the XIS spectrum vs. the best-fit continuum. The spectra have been rebinned for the purpose of viewing only.  Color representation is the same as in Fig.~\ref{spec}.     
\label{lineratio}}
\end{figure}

\begin{figure}
\includegraphics[scale=0.57,angle=270]{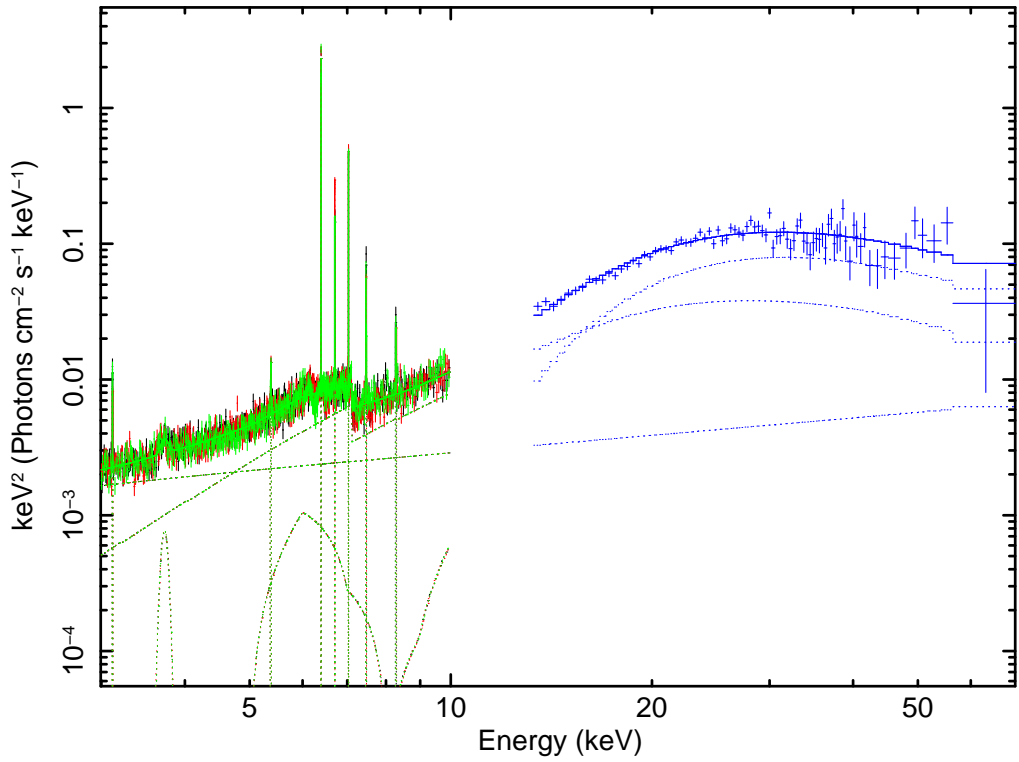}
\caption[models]{The unfolded spectrum between 3--70~keV band plotted in $\nu F_{\nu}$. Model 1 in the text has been used in unfolding the spectra. Color representations are the same as those in Fig.~\ref{spec}. 
\label{eeuf}}
\end{figure}

In Fig.~\ref{lineratio} we show the ratio of the XIS spectra and our baseline continuum model. The best-fit continuum model 
over 3--70~keV plus the significant emission lines are shown in Fig.~\ref{eeuf}. Emission lines detected in the 3--10~keV band 
are listed in Table~\ref{tab_lines}. We detected strong features that are consistent emission lines from neutral atoms of Fe and Ni:  
Fe K$\alpha$ (6.40 keV), Fe K$\beta$ (7.04 keV), and Ni K$\alpha$ (7.472). A weak line is detected at 6.71~keV, probably from 
He-like iron Fe K$\alpha$. The 7.04~keV line can also be a blend of the ``neutral'' Fe K$\beta$ and the 
H-like Fe K$\alpha$ (6.95~keV and 6.97~keV). We also found a significant line feature at $\sim 8.2$~keV ($> 3\sigma$). 
The feature is probably a blend of the H-like Fe K$\beta$ and He-like Fe K$\gamma$ emission lines, another indication of 
highly ionized gas. However, simply adding a high temperature gas component with APEC or MEKAL models in XSPEC cannot 
improve the fit and account for the flux of these lines. All except the 3.74~keV line are unresolved and thus modeled 
with $\delta$-functions to reduce the number of free parameters.  

\begin{figure}
\includegraphics[scale=0.57,angle=270]{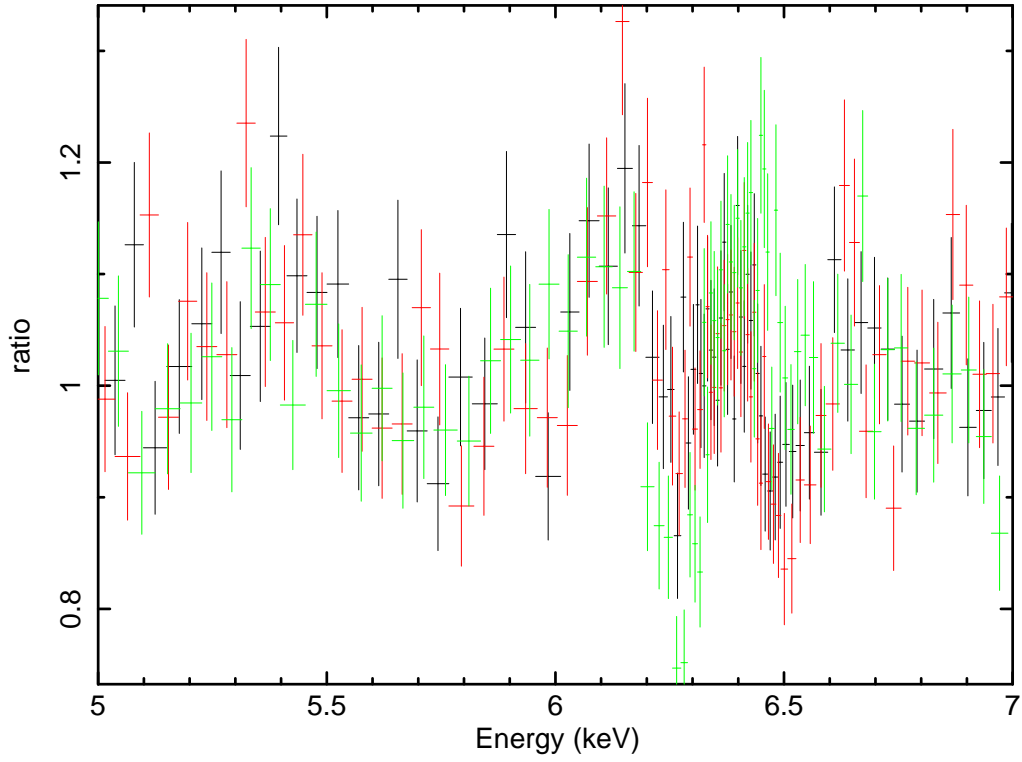}
\caption[models]{The ratio of the XIS spectrum and the model with the identified lines. The spectra have been rebinned for the purpose of viewing only.  Color representation is the same as in Fig.~\ref{spec}.     
\label{artefacts}}
\end{figure}

\begin{figure}
\includegraphics[scale=0.57,angle=270]{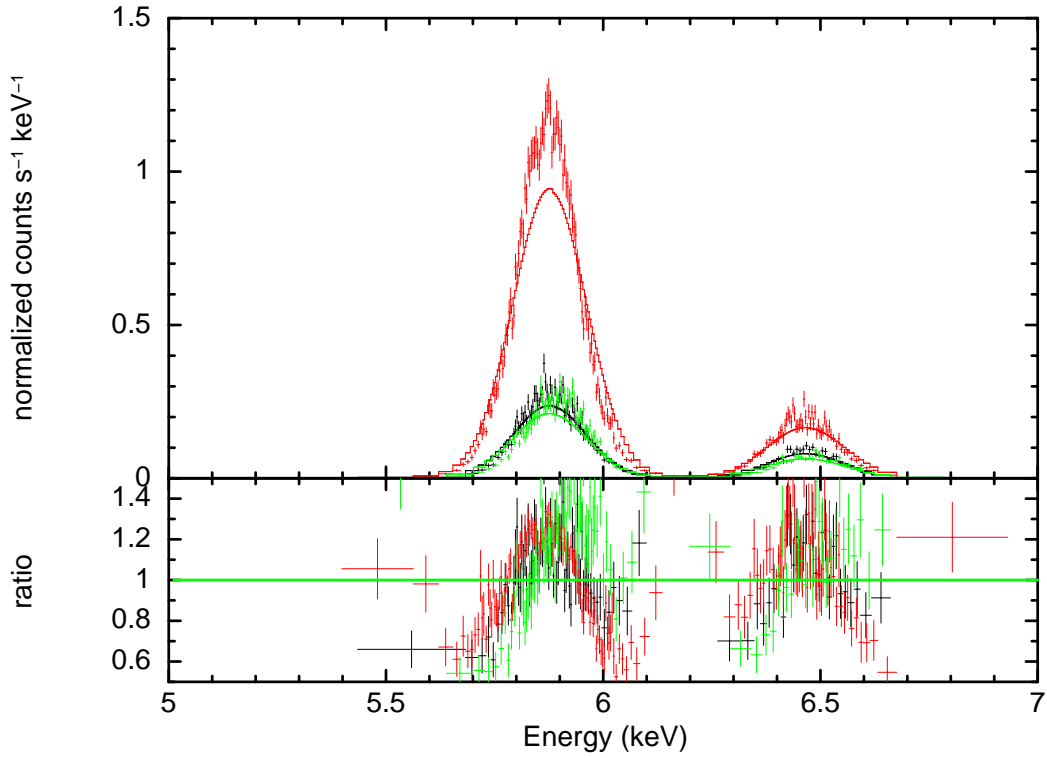}
\caption[cal]{Upper panel: The spectra of the calibration sources in XIS-0,2,3. The best-fit $\delta$-function models folded with the responses are
also shown. Lower panel: the ratio of the data and models in the upper panel. The color scheme is the same as Fig.~\ref{spec} 
\label{calib}}
\end{figure}

In Fig.~\ref{artefacts}, we show the ratio of XIS spectra between 5 and 7 keV and model 1 (\S~\ref{joint}) with only the identified lines shown 
in Table~\ref{tab_lines}. The spectra show no clear sign of Compton Shoulder at 6.23 keV. However, this is likely an artefact 
due to the inaccuracy of the calibration. To illustrate this, we show the XIS spectra of the calibration 
sources at the corners of the XIS detectors (Fig.~\ref{calib}). The Mn K$\alpha$ and K$\beta$ lines are modeled with 
Gaussians. The best-fit line width is consistent with 0, but the model line profiles are broader than those of the data. 
We measure the FWHM of the line model convolved with the response to be $\sim 190$~eV, while the FWHM of the  line is $\sim 180$~eV.  
For the strong  Fe K$\alpha$ line in the Circinus galaxy spectra, the broad wings of the response make it hard to properly 
fit the weak Compton shoulder only $\sim 0.2$~keV below the line energy. In Fig.~\ref{calib}, the broad response causes dips 
$\sim 0.1-0.2$~keV below and above the line energies in the plot of data to model ratio. The same is seen in Fig.~\ref{lineratio},
with a dip at $\sim 6.2-6.3$~keV, which coincident with the expected energy of the Compton shoulder. Broad wings are seen in the 
Fe K$\alpha$ line, which has also been observed in the {\it Chandra} observations (Sambruna et al. 2001). Part of the
low energy wing could attribute to the Compton shoulder.  A significant feature is detected at $\sim 5.4$~keV. The nature of the line, 
however, is not known.   

The iron abundance is found to be sub-solar in our models. This seem to disagree with the {\it XMM-Newton} results 
(Molendi et al 2003).  However, the Fe abundance estimate is sensitive to the soft power-law component in our model, which is
an over simplification of the spectra that includes emission from ULXs, the Thompson scattered nuclear emission, and the 
possible emission from hot gas. If we fit the 3--10~keV XIS spectra without the soft power-law, and fix the photon index
of the cold reflection component to 1.56, we find Fe abundance to be $\sim 1.35$ solar. Therefore, we can not conclude the
Fe abundance is actually different from the {\it XMM-Newton} results.    
It should be noted that if both reflection and absorption come from the torus, the abundance should be the same for the two spectral components. 
However, our absorption model only assumes solar abundance.  We estimate how Fe abundance affects the absorption column density. In Circinus galaxy, 
the ion edge  is dominated by the reflection component, and is determined by the XIS spectrum, while the absorption is determined by the PIN spectrum.   
We fit the PIN spectrum with XSPEC model zvarabs(cutoffpl), with the power-law index fixed to 1.56 and the high energy cut-off $E_C = 48.7$~keV.  
The abundances of all elements are fixed to the solar value except for  ion, which is fixed to the best-fit values found above. In this model, if Fe abundance 
is solar ($A_{Fe} = 1$), we found $N_H = 266 \pm 51$~\pcmsq; If  $A_{Fe} = 0.75(1.35)$, the best-fit $N_H = 318 \pm 60 (216 \pm 41)$~\pcmsq.  
In either case, the under or over estimate of  $N_H$ by assuming $A_{Fe} = 1$ is at $\sim 1\sigma$ level.  

If the flux in the 7.04~keV line is from Fe K$\beta$, the flux ratio of Fe~K$\beta$ and K$\alpha$ would be $0.20 \pm 0.02$. 
This value has not been corrected for the possible loss due to the low energy tails of the lines and the Compton Shoulder. 
The result is consistent with that from {\it Chandra} ($0.22 \pm 0.05$, \markcite{sambruna01}{Sambruna} {et~al.} (2001) and {\it BeppoSax} (0.191, 
\markcite{guainazzi99}{Guainazzi} {et~al.} (1999),  but appear to be significantly higher than that from {\it XMM-Newton} ($0.14 \pm 0.01$, Molendi et al. 2003)
and the expected value $\sim 0.16$ \markcite{basko78}({Basko} 1978).  
If the flux in the 7.04~keV is a blend of the ``neutral'' Fe K$\beta$ and the H-like Fe K$\alpha$, which we think likely based on 
our detection of lines that are consistent with the highly ionized species, the different line ratio
between the {\it XMM-Newton} and other observations may suggest variability in the ionization state of the gas, and could potentially
constrain the physical size of the ionization region. The detailed study of the ionized gas in Circinus galaxy is beyond the scope of
this paper.  
  
\section{Concluding remarks
\label{concl}}
Our main results from an X-ray observation of the Circinus galaxy with the {\it Suzaku} satellite are:

\begin{enumerate}
\item There is minimal contamination ($\pm 5\%$) of the 0.2--10~keV radiation by the periodic ULX CG X-1,
and less at higher energies.
\item Below 10~keV, we see a soft power-law ( that could come from a combination of extended emission, 
contaminating point sources in the galaxy, or the Thompson scattered nuclear emission) and a strong
cold reflection component (as observed by earlier satellites) from the nuclear emission.  
\item The spectrum above 20~keV is well described by a heavily absorbed Compton thick 
($\rm{N_H} = (4-5) \times 10^{24}$~\pcmsq) power-law model with a high energy cut-off 
($E_C \simeq 49$~keV). Based on the lack of variability in the PIN light curve, we suggest that the nucleus is obscured by a
torus with a large covering fraction, with the reflection component originating from the 
far side of the  torus. 
 \item The galaxy is detected in the 50--200~keV band at 2.5$\sigma$ level. This flux agrees well with 
 an extrapolation to the higher energies of the transmitted component seen in the 13--50~keV band. 
 \item We have detected  strong emission lines at energies consistent with the florescent lines of neutral 
 Fe K$\alpha$, K$\beta$, and  Ni K$\alpha$. These features are consistent with the reflection from cold 
 dense gas in the torus. However, we were not able to find the Compton shoulder in our XIS spectra.
 We point out this non-detection can be attribute to a calibration problem. 
 \item We also found emission features which are consistent with emission lines from He-like 
 Fe~K$\alpha$ (6.71~keV),  H-like Fe~K$\beta$ (8.2~keV). The emission line at 7.04~keV  is also consistent with the line energy of H-like Fe K$\alpha$.  These suggest that highly ionized gas
is present in the Circinus galaxy.  
 \end{enumerate}

\acknowledgements
Y. Yang would like to thank Dr Richard Mushotzky for discussions on the XIS spectra.  We thank Drs. Koji Mukai, Illana Harrus, and Kenji Hamaguchi at {\it Suzaku} GOF for help with issues related to data reduction and calibration. We thank the anonymous referee for suggestions that improve this paper. This work is supported by NASA through  LTSA grant NAG 513065 and {\it Suzaku} NMX06AI34G granted to the University of Maryland.  


\clearpage
\begin{deluxetable}{ccc}
\tablenum{1}
\tabletypesize{\footnotesize}
\tablewidth{0pt}
\tablecaption{Best-fit Continuum}
\tablecolumns{3} \tablehead{
\colhead{} &
\colhead{Model 1 } &
\colhead{Model 2}}  
\startdata
 $N_{H,T}$ ($10^{22}$ \pcmsq) & $463_{-23}^{+47}$     &  $470_{-32}^{+50}$ \\ 
 $A_T$\tablenotemark{a}     & $0.13_{-0.04}^{+0.03}$ & $0.14_{-0.05}^{+0.04}$\\ 
 $\Gamma_h$ & $1.56^{+0.09}_{-0.07}$  &  $1.58^{+0.07}_{-0.10}$  \\ 
 $E_C$ (keV) & $48.7$ (fixed) & $48.7$ (fixed) \\
$A_R$\tablenotemark{a} & $(2.2_{-0.3}^{+0.2}) \times 10^{-2}$ & $(2.4_{-0.4}^{+0.2}) \times 10^{-2}$  \\
$A_{Fe}$ (solar) & $0.75 \pm 0.09$ &  $0.74_{-0.06}^{+0.10}$ \\
$\Gamma_s$ & 1.54 (fixed) & $1.58=\Gamma_h$ \\
$A_s$\tablenotemark{a} & $1.1 \times 10^{-3}$ (fixed) & $ (1.0 \pm 0.3) \times 10^{-3}$  \\
$N_H$ ($10^{22}$ \pcmsq) & 0.38 (fixed) & 0.38 (fixed) \\
$\chisq/d.o.f$\tablenotemark{b} & 2066/1814 & 2067/1813 \\

\enddata
\tablenotetext{a}{unit: $\rm{ph \thinspace keV^{-1} \thinspace cm^{-2}
    \thinspace s^{-1}}$ at 1 keV.} 
\tablenotetext{b}{The $\chisq$ are obtained with added emission lines listed in Table~\ref{tab_lines}.}
\label{tab_cont}
\end{deluxetable}

\begin{deluxetable}{ccccc}
\tablenum{2}
\tabletypesize{\footnotesize}
\tablewidth{0pt}
\tablecaption{Emission lines}
\tablecolumns{5} \tablehead{
\colhead{Energy (keV)} &
\colhead{Identification } &
\colhead{flux \tablenotemark{a}} &
\colhead{Gaussian $\sigma$ (keV)\tablenotemark{b}} & 
\colhead{EW (keV)}}
\startdata
$3.12 \pm 0.02   $  &  Ar XVII                      &  $7.2_{-1.7}^{+2.0} \times 10^{-6}$ &  0 & 0.032\\
$3.75 \pm 0.05$     & Ca XIX, Ar XVII           &  $7.81_{-0.06}^{+0.08}  \times 10^{-6}$ & $(7.3 \pm 0.05) \times 10^{-2}$ & 0.040\\
$5.38 \pm 0.04$     & unkown                & $4.22 \times 10^{-6}$                 & 0 & 0.022\\
$6.405 \pm 0.001$  & Fe K$\alpha$ (II-XVII) & $(3.45 \pm 0.04) \times 10^{-4}$ & 0 & 1.49  \\
$6.716 \pm 0.006$  & Fe K$\alpha$ ($>$XVII) & $(4.7 \pm 0.02) \times 10^{-5}$ & 0 & 0.049  \\
$7.040 \pm 0.004$  & Fe K$\beta$                & $(6.95 \pm 0.02) \times 10^{-5}$ & 0 & 0.126  \\
$7.48 \pm 0.01 $   & Ni K$\alpha$               & $(1.66 \pm 0.02) \times 10^{-5}$ & 0 & 0.081  \\
$8.28 \pm 0.04 $   & Fe K$\beta$ (XXVI)+ K$\gamma$ (XXV) & $6.01 \times 10^{-6}$ & 0 & 0.058 \\     

\enddata
\tablenotetext{a}{Unit: ph~cm$^{-2}$~s$^{-1}$}
\tablenotetext{b}{Most of the lines have $\sigma$ consistent with 0, so their width is set to 0 to reduce the number of free parameters.}
\label{tab_lines}
\end{deluxetable}

\clearpage

%

\vfil\eject\clearpage
\vfil\eject\clearpage
\enlargethispage*{2000pt}
\vfil\eject\clearpage
\vfil\eject\clearpage

\end{document}